\documentclass[fleqn,usenatbib,useAMS]{mnras}

\usepackage[T1]{fontenc}
\usepackage{ae,aecompl}

\usepackage{graphicx}	
\usepackage{amsmath}	
\usepackage{amssymb}	

\title[The very slow rotation of HD\,54879]{
The very slow rotation of the magnetic O9.7\,V star HD\,54879
}

\author[Hubrig et al.\ 2019]{
S.~Hubrig$^{1}$\thanks{E-mail: shubrig@aip.de}
S.~P.~J\"arvinen$^{1}$,
M.~Sch\"oller$^{2}$,
and C.~A.~Hummel$^{2}$
\\
$^{1}$Leibniz-Institut f\"ur Astrophysik Potsdam (AIP), An der Sternwarte~16, 14482~Potsdam, Germany\\
$^{2}$European Southern Observatory, Karl-Schwarzschild-Str.~2, 85748 Garching, Germany\\
}
\date{Accepted XXX. Received YYY; in original form ZZZ}

\pubyear{2019}

\begin{document}
\label{firstpage}
\pagerange{\pageref{firstpage}--\pageref{lastpage}}
\maketitle

\begin{abstract}
The first FORS\,2 spectropolarimetric observation of the longitudinal
magnetic field of HD\,54879 of the order of $-600$\,G with a lower limit of the dipole
strength of $\sim$~2\,kG dates back to 2014. Since then observations showed a gradual decrease of the
absolute value of the mean longitudinal magnetic field.
In the course of the most recent monitoring of HD\,54879 using FORS\,2 spectropolarimetric 
observations from 2017 October to 2018 February, a longitudinal magnetic field strength change from
about $-300$\,G down to about $-90$\,G was reported.
A sudden increase of the
absolute value of the mean longitudinal magnetic field
and an accompanying spectral variability was detected on 2018 February 17.
New FORS\,2 spectropolarimetric data obtained from 2018 December to 2019 February
confirm the very slow magnetic field
variability, with the field decreasing from about 150\,G to $-$100\,G over two months.
Such a slow magnetic field variability, related to the extremely slow rotation of HD\,54879, is also confirmed using high-resolution
HARPS\-pol and ESPaDOnS spectropolarimetry. 
The re-analysis of the FORS\,2 polarimetric spectra from 2018 February
indicates that the previously reported field increase
and the change of the spectral appearance
was caused by improper spectra extraction and wavelength calibration using observations obtained at
an insufficient
signal-to-noise ratio. The presented properties of HD\,54879 are discussed in the context of the Of?p 
spectral classification.
\end{abstract}




\section{Introduction}
\label{sec:intro}

Only few massive O-type stars possess strong large-scale organized
magnetic fields. One of them, the O9.7\,V star HD\,54879
was detected as magnetic by \citet{Castro2015}
using the FOcal Reducer low dispersion Spectrograph (FORS\,2; \citealt{Appenzeller1998})
and the High Accuracy Radial 
velocity Planet Searcher in polarimetric mode (HARPS\-pol; \citealt{Snik2008}).
The authors reported the presence of a $-600$\,G longitudinal magnetic
field with a lower limit of the dipole strength of $\sim2$\,kG. 
Numerical MHD simulations describing the interaction of the magnetic field of this star with its wind were
recently discussed by \citet{Hubrig2019}.
In the same work, \citet{Hubrig2019} presented 26 new FORS\,2 spectropolarimetric observations of HD\,54879  
obtained from 2017 October~4 to 2018 February~21 showing a change of the
mean longitudinal magnetic field
from about $-300$\,G down to about $-90$\,G. 
The authors also reported on a sudden, short-term increase of the
absolute value of the mean longitudinal magnetic field
on the night of 2018 February~17,
measuring a longitudinal magnetic field of $-833$\,G. The inspection of the FORS\,2 spectrum acquired during 
that night also indicated a change in spectral appearance and a decrease of the radial velocity by several
10\,km\,s$^{-1}$.

Since such an unusual behaviour was not observed for other massive magnetic stars,
we decided to monitor the magnetic field variability again
and obtained data between 2018 December and 2019 February.
In the following, we report on our results obtained from these most recent FORS\,2 spectropolarimetric observations,
indicating a very slow change of the magnetic field related to slow stellar rotation.
High-resolution spectropolarimetric observations confirm the results of the magnetic field measurements made
using low-resolution FORS\,2 spectropolarimetric observations.
Our new results are also discussed in the context of the previous detection of a short-term field increase
and the change of spectral appearance. 


\section{Observations and magnetic field analysis}
\label{sec:obs}

24 new spectropolarimetric observations of HD\,54879  with FORS\,2 were obtained between
2018 December~16 and 2019 February~15 in the framework of our programme 0102.D-0285 executed in service mode
at the ESO VLT. Among them, five observations were obtained outside the weather 
specifications and appear underexposed  with signal-to-noise ratios ($S/N$) less than 1300, whereas one observation
yielded  saturated spectra.
In the following, we will discuss the 18 unsaturated high $S/N$ spectra,
while we discuss the five lower $S/N$ spectra in more detail in Sect.~\ref{sec:lowSNR}.

All spectra were observed with the GRISM 600B and the narrowest available slit width of 0$\farcs$4 to obtain
a spectral resolving power of $R\approx2000$. The observed spectral range from
3250 to 6215\,\AA{} includes all Balmer lines, apart from H$\alpha$, and
numerous helium lines. Further, in our observations, we used a non-standard
readout mode with low gain (200kHz,1$\times$1,low), which provides a broader
dynamic range, hence allowing us to reach a higher $S/N$
in the individual spectra. 
The extraction of the parallel and perpendicular beams on the FORS\,2 raw data
was carried out using a pipeline written in the MIDAS environment
by T.~Szeifert, (in the following called the MIDAS pipeline).
More details on the observational and reduction methods can be found in \citet{Hubrig2019} 
and references therein.

\begin{table}
\centering
\caption{Longitudinal magnetic field measurements of HD\,54879 using
low resolution FORS\,2 spectropolarimetric observations. 
The $S/N$ is measured at 4800\,\AA{}.
}
\label{tab:FORS}
\begin{tabular}{ccr@{$\pm$}lr@{$\pm$}lr@{$\pm$}lc}
\hline
\multicolumn{1}{c}{MJD} &
\multicolumn{1}{c}{$S/N$} &
\multicolumn{2}{c}{$\left<B_{\rm z}\right>_{\rm all}$} &
\multicolumn{2}{c}{$\left<B_{\rm z}\right>_{\rm hyd}$} &
\multicolumn{2}{c}{$\left<B_{\rm z}\right>_{\rm N}$} \\
& 
&
\multicolumn{2}{c}{(G)} &
\multicolumn{2}{c}{(G)} &
\multicolumn{2}{c}{(G)}
\\
\hline
58468.1062    & 1740 & 153 & 91 & 216 & 169 & 12 & 89  \\
58474.3364    & 2210 &  82 & 63 &  32 & 140 &34 & 65  \\
58478.3144    & 2910 &  5  & 59 & 135 & 123 &4  & 66 \\
58479.2787    & 2220 &  $-$71 & 88& 40 & 137 &$-$44& 80 \\
58480.0776    & 2340 & $-$32 & 102& 24  & 148 &19 & 112 \\
58482.2096    & 2390 & 42  & 65 & $-$49  & 97  &$-$6  & 63 \\
58483.0487    & 1520 & $-$20  & 107 & $-$101 & 170 &78 & 103  \\
58484.1973    & 2040 & 38  & 81 & 76  & 123 &  15  & 79 \\
58488.1147    & 2620 & $-$55  & 73 & $-$21  & 128 & 74   & 78 \\
58489.3414    & 1870 & $-$48  & 92 & $-$24  & 129 & $-$12& 88 \\
58493.3380    & 2300 &  $-$65 & 59 & 81  &  99 &   23 & 63 \\
58496.0606    & 2110 &  $-$22  & 81 &  $-$89  &  105 & $-$78& 78 \\
58511.0924    & 2280 &  67  & 80 & 14 &  107 & $-$52& 83 \\
58524.0605    & 2440 & 45   & 67 & $-$19  &  98 & 16& 65 \\
58525.1362    & 1980 & 125 & 103 & 61 & 138 & $-$41& 99 \\
58526.2129    & 2080 & 50 & 82 & 117 & 132 & 10& 78 \\
58527.1349    & 2210 & $-$102 & 88 & 39 & 138 & $-$35& 89 \\
58529.1940    & 1960 & $-$107 & 99 & $-$292 & 160 & $-$68& 103 \\
\hline
\end{tabular}
\end{table}

The results of the measurements are listed in Table~\ref{tab:FORS},
where the first two columns give the modified Julian dates
(MJDs) for the middle of the exposure and the $S/N$ values of the spectra.
The results of our magnetic field measurements,
those for the entire spectrum and those using only the hydrogen lines, are presented 
in Columns~3 and 4, followed by the measurements using all lines
in the null spectra. 
Null spectra $N$ are calculated as pairwise differences from all available
Stokes~$V$ profiles so that the real polarization signal should cancel out. 

\begin{figure}
 \centering 
\includegraphics[width=0.80\columnwidth]{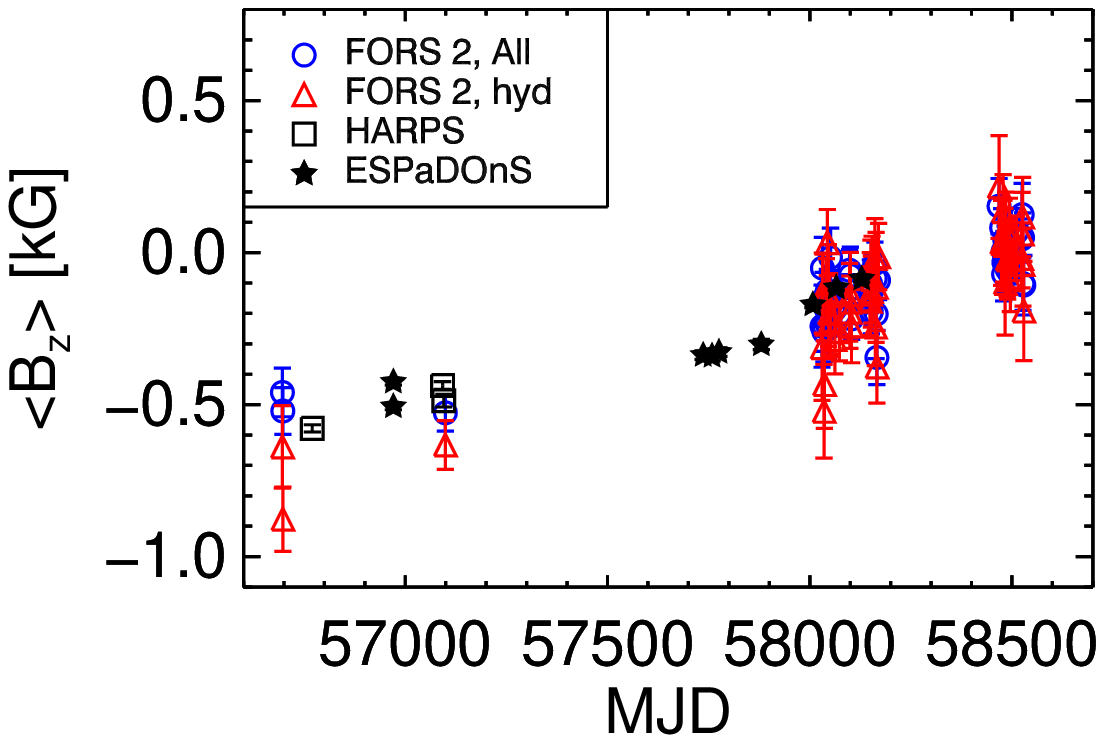}
\includegraphics[width=0.80\columnwidth]{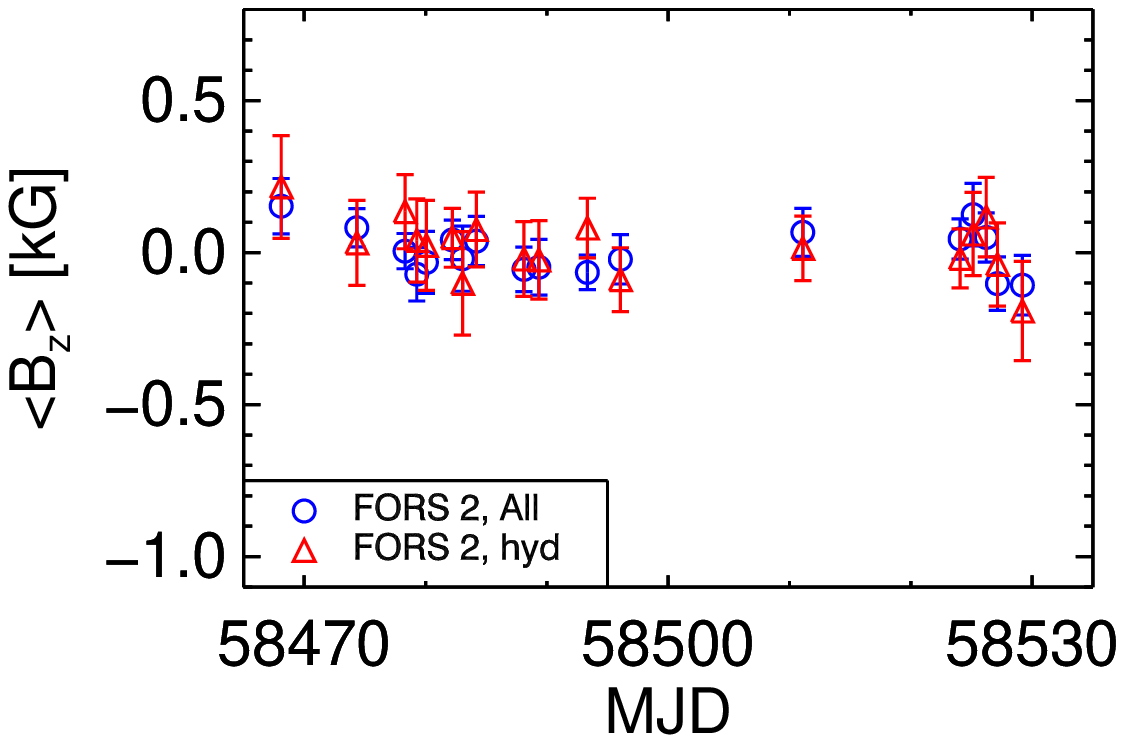}
                \caption{
{\it Top panel:} Distribution of the mean longitudinal magnetic field values of
   HD\,54879 using the entire spectrum (open blue circles) and those using only the hydrogen lines 
(open red triangles) as a function of MJD between the years 2014 and 2019. Open black squares and filled stars indicate
high-resolution spectropolarimetric observations with HARPS and ESPaDOnS, respectively. 
{\it Bottom panel:}  Distribution of the mean longitudinal 
magnetic field values of HD\,54879 measured with FORS\,2 as a function of MJD between 2018 December~16 and 2019 February~15.
         }
   \label{fig:Bevol}
\end{figure}

The distribution of the mean longitudinal
magnetic field values as a function of MJD is
presented in Fig.~\ref{fig:Bevol}.
In the top panel of this figure, we show all FORS\,2 longitudinal magnetic field measurements
acquired between 2014 February 8 and 2019 February 20. The most recent measurements obtained 
from December 2018 to February 2019 are presented in the bottom panel.  
While the few observations from 2014 and 2015 indicated
a mean longitudinal magnetic field
of the order of $-$500\,G to $-$900\,G, we observe in the last 
years a significantly weaker longitudinal magnetic field with
values 
between $-300$\,G and $+150$\,G. 
The strongest longitudinal magnetic field of positive polarity of 150\,G
was measured on the night 2018 December~16.
After this date the longitudinal magnetic field is gradually decreasing,
reaching
a value
of about $-$100\,G on 2019 February~15.
These measurements together with those presented
by \citet{Hubrig2019} indicate that the rotation period of HD\,54879 is very long, at least several years, and that 
from 2017 October to 2019 February we are observing the star 
at rotational phases with the best visibility of the magnetic equator.
A small scatter in the measurements of the magnetic field
is frequently observed in massive O-type stars
and is most likely due to contamination by the immediate stellar environment
with a denser cooling disc,
confined to the magnetic equatorial plane (e.g.\ \citealt{Hubrig2015,Shultz2017}).
However, the scatter we see in the measurements presented in Fig.~\ref{fig:Bevol}
is within the uncertainties represented by the error bars.

To check the consistency of the low-resolution FORS\,2 magnetic field measurements with measurements using
high-resolution spectropolarimeters,
we overplotted in Fig.~\ref{fig:Bevol} these FORS\,2 measurements
with high-resolution longitudinal magnetic field measurements obtained from archival data.
We discuss these observations in the following section.

\section{High-resolution spectropolarimetry of HD\,54879}
\label{sec:esp}

\begin{figure}
 \centering 
\includegraphics[width=0.80\columnwidth]{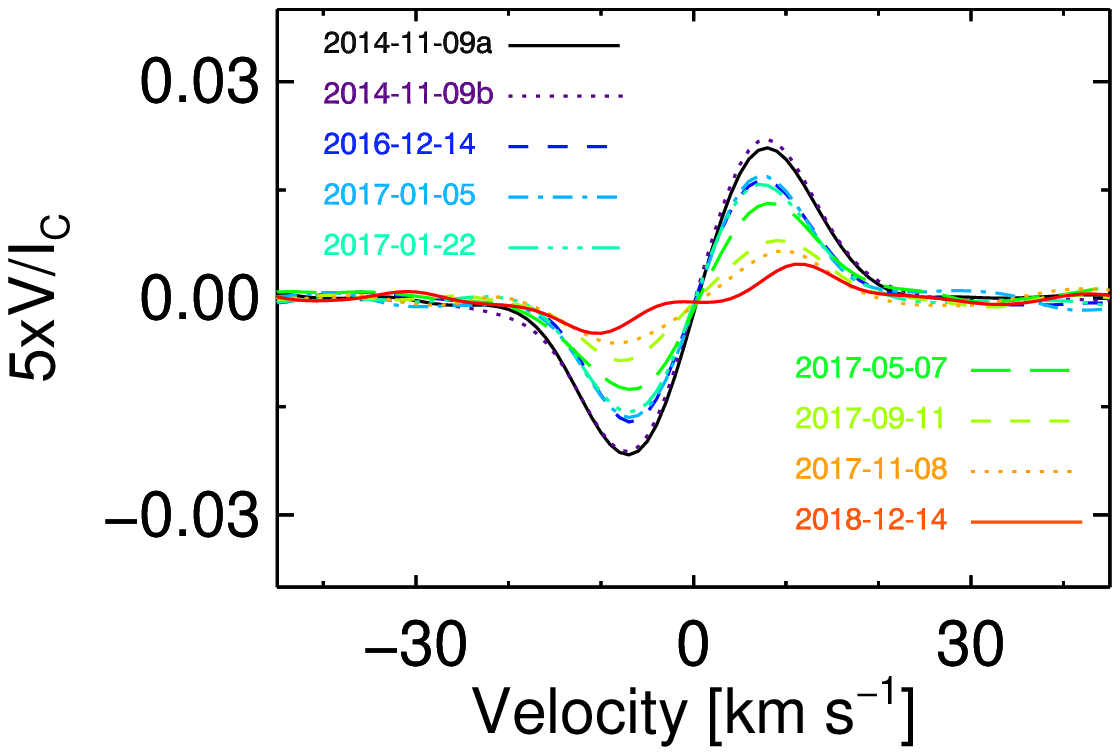}
\includegraphics[width=0.80\columnwidth]{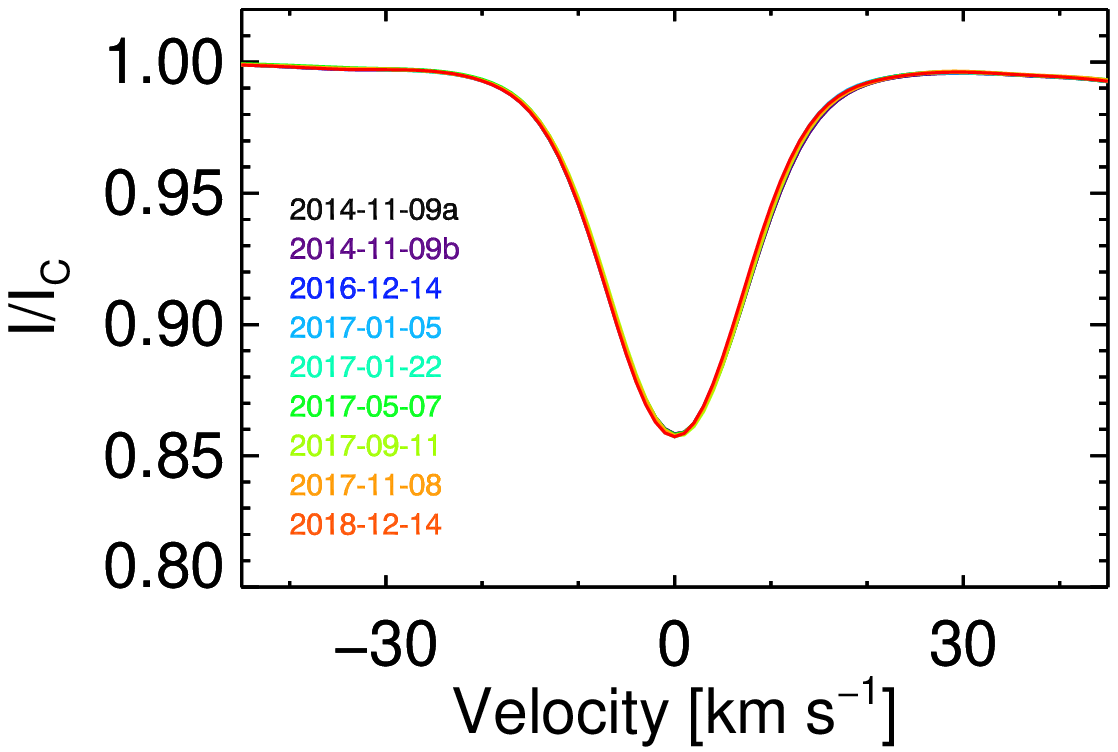}
                \caption{
LSD Stokes~$V$ and Stokes~$I$ profiles computed using the ESPaDOnS spectra 
obtained from 2014 November~9 to 2018 January~10.
         }
   \label{fig:esp1}
\end{figure}

\begin{table}
\centering
\caption{
The mean longitudinal magnetic field
measured for HD\,54879 from
  ESPaDOnS polarimetric spectra in two different ways,
using all lines apart from the hydrogen lines and using exclusively metal lines.
  The $S/N$ values are measured at 6500\,\AA{}.
}
\label{tab:espa}
\begin{tabular}{cccr@{$\pm$}lr@{$\pm$}l}
\hline
\multicolumn{1}{c}{MJD} &
\multicolumn{1}{c}{$t_{\rm exp}$} &
\multicolumn{1}{c}{$S/N$} &
\multicolumn{2}{c}{$\left<B_{\rm z}\right>_{\rm all}$} &
\multicolumn{2}{c}{$\left<B_{\rm z}\right>_{\rm met}$}\\
&
\multicolumn{1}{c}{(min)} &
&
\multicolumn{2}{c}{(G)} &
\multicolumn{2}{c}{(G)} 
\\
\hline
56970.56& 56  & 407   & $-$327 & 10 & $-$425 & 11 \\
56970.60& 56  & 403   & $-$386 & 10 & $-$506 & 11 \\
57736.48& 32  & 336   & $-$267 & 14 & $-$336 & 14 \\
57758.37& 44  & 361   & $-$280 & 13 & $-$339 & 16 \\ 
57775.48& 56  & 340   & $-$230 & 13 & $-$325 & 14 \\
57880.23& 48  & 343   & $-$239 & 12 & $-$301 & 17 \\
58007.62& 59  & 370   & $-$135 &  8 & $-$170 & 10 \\
58065.53& 59  & 424   & $-$86  &  9 & $-$114 & 10 \\
58128.41& 59  & 341   & $-$61  &  8 & $-$85  & 11 \\
\hline
\end{tabular}
\end{table}

As already reported by \citet{Jarvinen2017}, three spectropolarimetric observations of HD\,54879 were
obtained with the HARPS polarimeter attached to ESO's 3.6~m telescope (La~Silla, Chile)
at a spectral resolution of about 115\,000
on 2014 April~22, and on 2015 March~11 and March~14.
The published
values of the mean longitudinal magnetic field
showed a change from $-$578\,G in 2014 to  $-$487\,G in 2015.
Recently, nine ESPaDOnS (Echelle SpectroPolarimetric Device for the Observation of Stars; \citealt{Donati2006})
spectra with a spectral resolution of 65\,000
obtained between 2014 November and 2018 January
became publicly available in the CFHT archive.
Among them, two observations were obtained on the same night on 2014 November~9.
To measure the longitudinal magnetic field, we employed 
the least-squares deconvolution (LSD) technique, allowing
us to achieve a much higher $S/N$ in the polarimetric spectra.  The LSD technique combines line 
profiles (assumed to be identical) centred on the position of the 
individual lines and scaled according to the line strength and 
sensitivity to a magnetic field.
The details of
this technique and of the calculation of the Stokes~$I$ and Stokes~$V$
parameters can be found in the work of \citet{Donati1997}. 
The line masks employed in the measurements of the longitudinal magnetic fields, one with 127 lines 
including the He and the metal lines and the second one with 113 metal lines, 
were constructed using the Vienna Atomic Line Database
\citep[VALD3;][]{kupka2011} and the stellar parameters $T_{\rm eff}=30.5$\,kK  and $\log\,g=4.0$ reported by \citet{Shenar2017}.
The calculated LSD Stokes~$I$ and Stokes~$V$ profiles for each observing epoch are presented in Fig.~\ref{fig:esp1}.
The measurements using both line masks are listed in Table~\ref{tab:espa} along with the modified Julian dates
for the middle of the exposure, the exposure times, and the $S/N$ values of the spectra. 
In all cases the false alarm probability was less than $10^{-10}$.
Notably, the distribution of data points obtained for the high-resolution spectropolarimetric observations 
in Fig.~\ref{fig:Bevol} fits the low-resolution FORS\,2 measurements
very well, indicating that the measurements using low resolution spectropolarimetry
are fully consistent with those made with other instruments.

\section{Rotation period and short-term variability}
\label{sec:per}

The magnetic field measurements presented in Fig.~\ref{fig:Bevol} at MJDs between 56696 and 58529 suggest that
the rotation period of HD\,54879 is very long.
Assuming that the negative field extremum reaches
a value
of $-500$\,G to $-900$\,G, measured in 2014 February,
and not having reached yet this value by 2019 February,
it is very likely that the rotation cycle is longer than five years.    

 \begin{figure}
\centering 
\includegraphics[width=0.80\columnwidth]{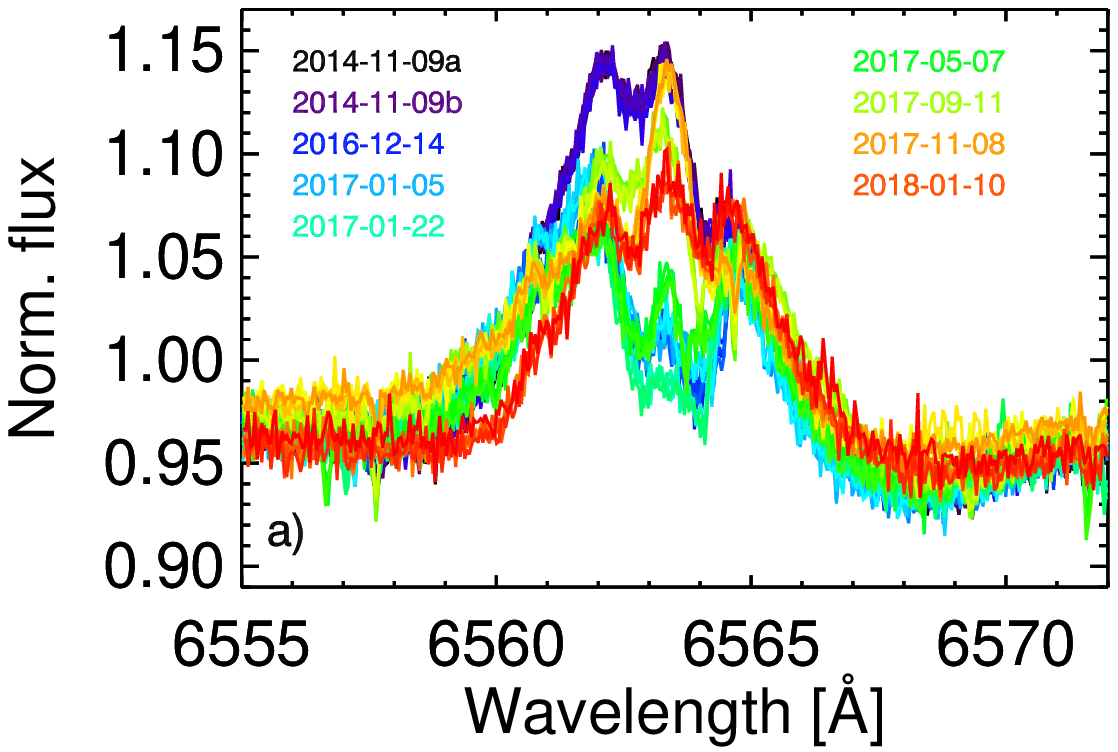}
\includegraphics[width=0.80\columnwidth]{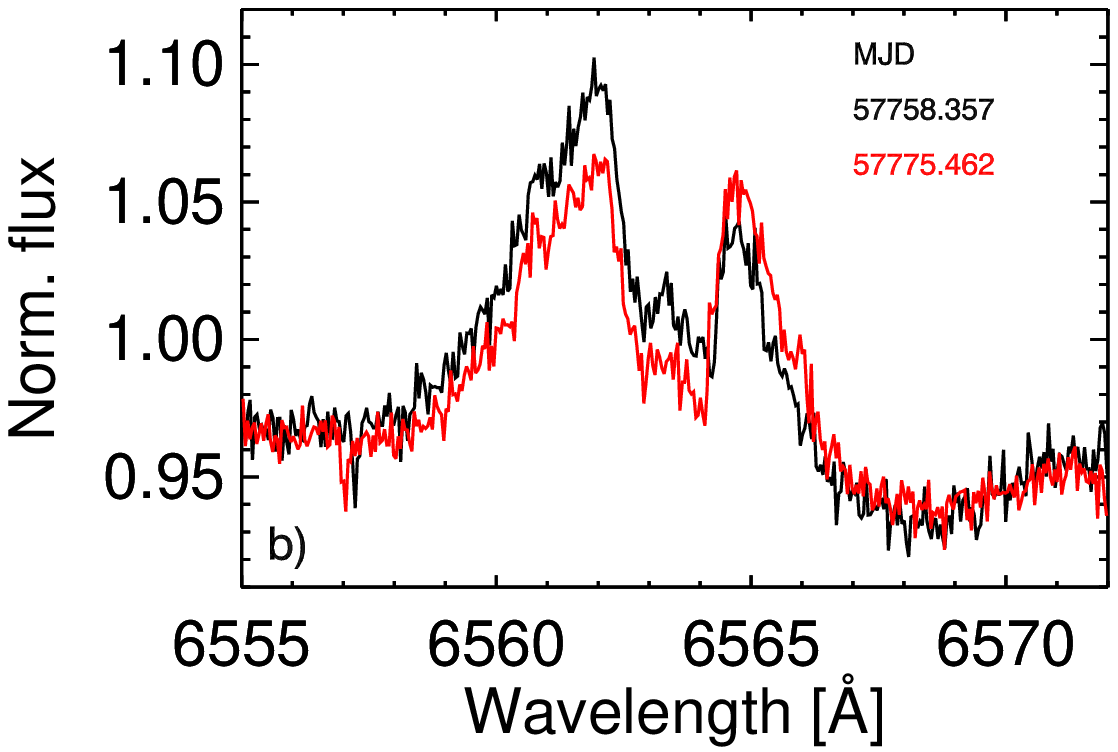}
\includegraphics[width=0.80\columnwidth]{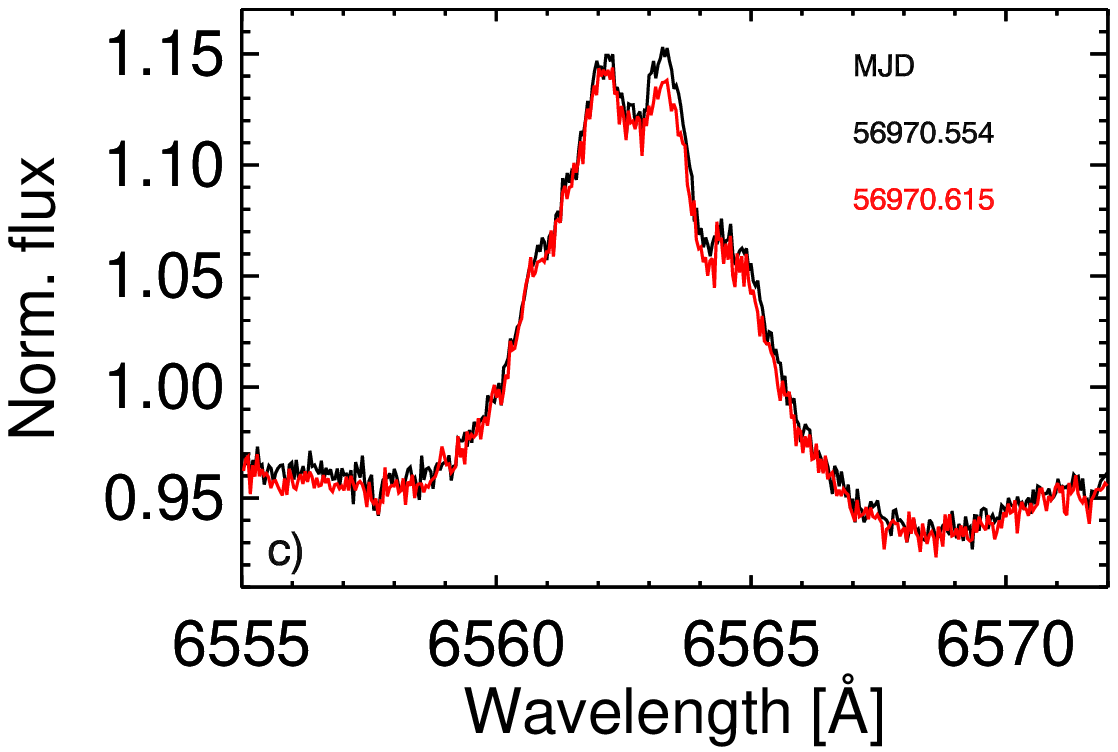}
\includegraphics[width=0.80\columnwidth]{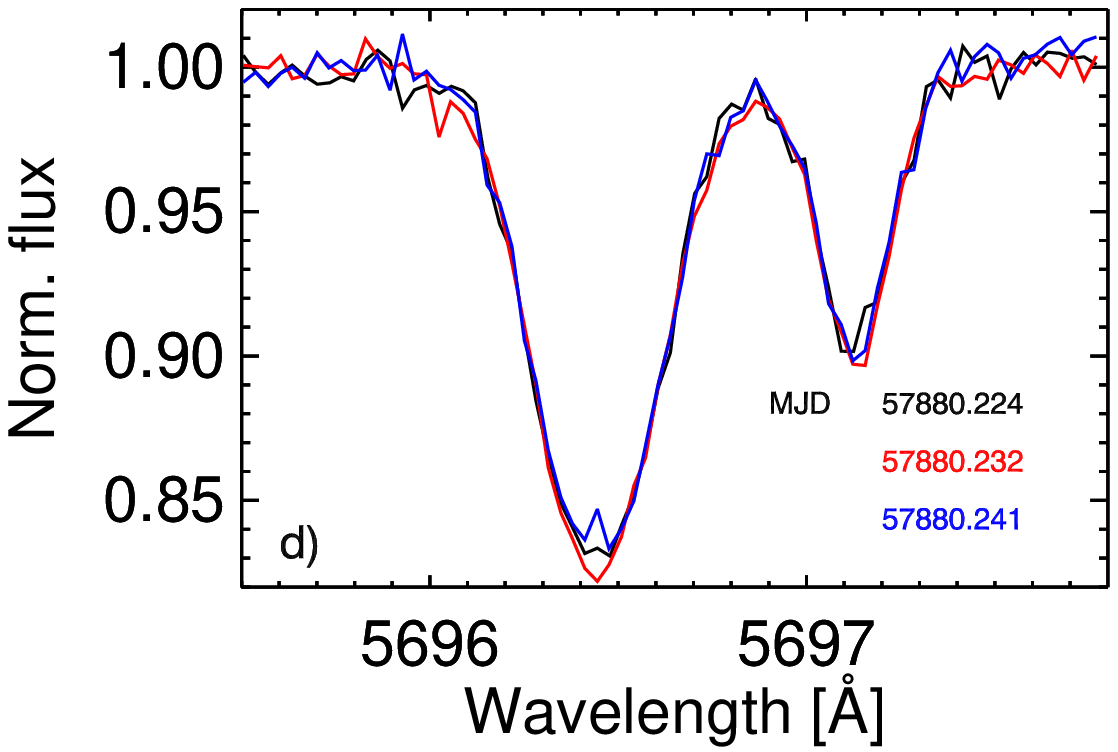}
               \caption{
Stokes~$I$ profiles recorded in ESPaDOnS spectra showing variability of the H$\alpha$ line profile
on short- and long-term scales.
The small changes found in the line cores of the \ion{C}{iii}~5696 and 
\ion{Si}{iii}~5697 lines on a time scale of a few minutes and displayed in panel d
are of the same order as the noise in the continuum and need to be verified with higher $S/N$ data.
         }
   \label{fig:esp2}
\end{figure}

Additional evidence for a very long rotation period follows from the consideration of the variability of the
H$\alpha$ line profiles. 
Previous studies of magnetic massive stars revealed a correlation between the
absolute value of the mean longitudinal magnetic field
and the strength of the H$\alpha$ emission in the sense that the strongest H$\alpha$ emission appears at phases of 
maximum
absolute value of the mean longitudinal magnetic field \citep{Sundqvist2012}.
Also the light curves in the
visible and the X-ray emission strengths usually display a positive correlation with the
absolute value of the mean longitudinal magnetic field.
In Fig.~\ref{fig:esp2}a we present profiles of  H$\alpha$ emission observed in the high-resolution
ESPaDOnS spectra. The shape of the H$\alpha$ emission observed in these spectra 
obtained from 2014 November to 2018 January is extremely variable,
changing between a triple-peak and a double-peak emission-line profile. Such a strongly variable
shape of the H$\alpha$ emission reflects a composite structure
of the surrounding circumstellar material.
We see that the lowest emission contributions with the deepest V-shaped central depression
have appeared in 2017 January. This is in agreement with the 
H$\alpha$ equivalent width (EW) measurements presented in Table A1 in the work of \citet{Shenar2017}, who detect
the lowest emission EW in 2017 February. The observations of the H$\alpha$ profile variability presented by these authors
in their Figure~13 go back to 2009. However, none of the displayed H$\alpha$ profiles shows such a low emission level
as observed in 2017 January in the ESPaDOnS spectra.
Since other massive magnetic stars show a variability of the H$\alpha$
emission strength with the stellar rotation period, the fact that the lowest H$\alpha$ emission profile has been observed only once 
since 2009 may suggest that the rotation period of HD\,54879 is longer that 9\,yr. 

In Fig.~\ref{fig:esp2}b and \ref{fig:esp2}c we show the variablity of H$\alpha$ on different, much shorter time scales.
The two profiles overplotted in Fig.~\ref{fig:esp2}b were observed in ESPaDOnS spectra on a time scale of 17\,d, whereas those
presented in Fig.~\ref{fig:esp2}c have a time lapse of only 88\,min.
A small variability in the line cores of \ion{C}{iii}~5696 and \ion{Si}{iii}~5697
on a time scale of a few minutes can be seen in Fig.~\ref{fig:esp2}d,
which is however on the same scale as the noise in the continuum
and will need higher $S/N$ data for verification.
The observed short-term spectral variability is not expected to be caused by changes in the star's stellar parameters,
but might be related to the wind or immediate environment of the star, including a denser cooling disc confined to the
magnetic equatorial plane (e.g.\ \citealt{Martins2012}). In absence of sufficient centrifugal support due to the slow rotation,
material accumulated in the disc and located below the corotation radius falls back onto the stellar
surface.   

It is of interest that the typical Of?p star HD\,108 with a rotation period of several decades
shows \ion{He}{i} and Balmer line variability on time-scales of days,
similar to the observed short-term variability of HD\,54879.
A short-term variability on the time scale of hours was reported for another Of?p star,
CPD\,$-$28$^{\circ}$\,2561, by \citet{Hubrig2015}.
 The variability domain of minutes or tens of minutes in magnetic O-type stars
remains however until now unexplored due to the lack of suitable spectroscopic and photometric time-series.

\begin{figure}
 \centering 
        \includegraphics[width=0.99\columnwidth]{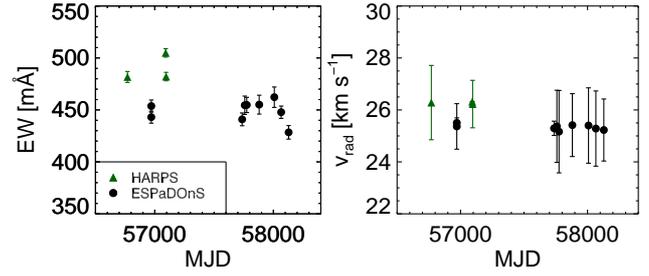}
                \caption{
Equivalent widths (left panel) and radial velocities (right panel) of the \ion{He}{ii}~4686 line measured in the 
high-resolution HARPS and ESPaDOnS spectra obtained from 2014 April to 2018 January.
}
   \label{fig:hrv}
\end{figure}

Apart from H$\alpha$, also the \ion{He}{ii}~4686 line is a very sensitive stellar wind indicator and can be used to
study the variability in high-resolution HARPS\-pol and ESPaDOnS spectra.
In Fig.~\ref{fig:hrv} we present our measurements of equivalent widths and radial velocities of this line in 
spectra obtained from 2014 October to 2018 January. The equivalent widths (EWs) measured on the HARPS\-pol spectra
appear larger than those measured in the ESPaDOnS spectra, probably due to the much higher spectral resolution of HARPS\-pol. 
In spite of the large dispersion of measurements in the spectra obtained in the last years, it appears quite possible 
that the distribution of the data points for the ESPaDOnS spectra indicates a small decrease of the 
strength of the \ion{He}{ii}~4686 line in 2018. 
The results of the radial velocity measurements presented in this figure on the right side show a 
slightly decreasing trend, which however is of the order of the measurement accuracies.

\begin{figure}
 \centering 
        \includegraphics[width=0.99\columnwidth]{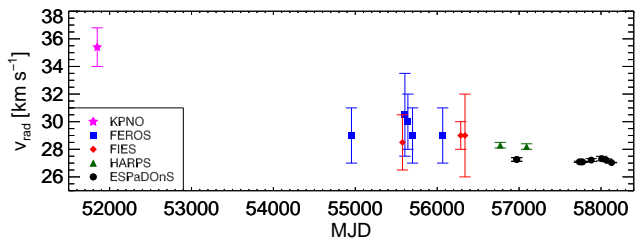}
                \caption{
Radial velocity measurements of HD\,54879 between 2000 December and 2018 January compiled from the literature 
and supplemented by the most recent HARPS\-pol and ESPaDOnS observations.}
   \label{fig:rv}
\end{figure}

The first radial velocity measurement reported in the literature, $v_{\rm rad}=15.6\pm1.4$\,km\,s$^{-1}$ by \citet{Neubauer1943},
was followed by the work of \citet{Boyajian2007}, who reported $v_{\rm rad}=35.4\pm1.4$\,km\,s$^{-1}$.
\citet{Castro2015} compared in their Table~4 the radial velocity measurement of $29.5\pm1.0$\,km\,s$^{-1}$ from one HARPS spectrum
with other measurements in the literature and concluded that HD\,54879 could be a member of a long-period binary system.
Our own measurements using the ESPaDOnS Stokes~$I$ spectra and employing the LSD technique indicate that the radial velocity
slightly decreased to 
$v_{\rm rad}=27.0\pm0.1$\,km\,s$^{-1}$ measured on the most recent ESPaDOnS spectra obtained in 2018 January.
In Fig.~\ref{fig:rv} we present the compiled literature measurements complemented by the three HARPS
and nine most recent ESPaDOnS observations.
A very slow but gradual decrease in radial velocity can probably be considered as real in
view of the high measurement accuracy reached in high-resolution spectropolarimetric observations using the LSD technique and
indicates that HD\,54879 could be a member of a binary system with a very long orbital period. Obviously, future 
monitoring of HD\,54879  with high-resolution spectroscopy is necessary to confirm the observed 
decrease in EWs and radial velocities.

\begin{figure}
 \centering 
        \includegraphics[width=0.99\columnwidth]{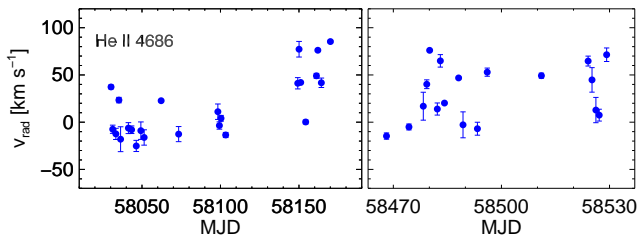}
                 \caption{
{\it Left:} Radial velocities for the \ion{He}{ii}~4686 line measured in the FORS\,2 spectra obtained 
from 2017 October~4 to 2018 February~21.
{\it Right:} Radial velocities for the \ion{He}{ii}~4686 line measured in the FORS\,2 spectra obtained 
from 2018 December~16 to 2019 February~15.
}
   \label{fig:RVprev}
\end{figure}

We also tested whether it is possible to use the low resolution of FORS\,2 observations to study the
variability of the 
\ion{He}{ii}~4686 line over the time interval from 2017 October~4 to 2019 February~15.
The radial velocity changes for the \ion{He}{ii}~4686 line in FORS\,2  spectra acquired between 
2017 October~4 and 2018 February~21 are presented in Fig.~\ref{fig:RVprev} on the left side and those measured on the 
spectra obtained between 2018 December 16 and 2019 February 15 on the right side of the same figure. 
It is obvious that the 
scatter of the data points presented in both panels is too big to allow us to make any conclusion
on the variability of the radial velocities.
This result is in line with the previous inconclusive search of 
periodicity by \citet{Hubrig2019} using FORS\,2 radial velocities. 

\begin{figure}
 \centering 
\includegraphics[width=0.99\columnwidth]{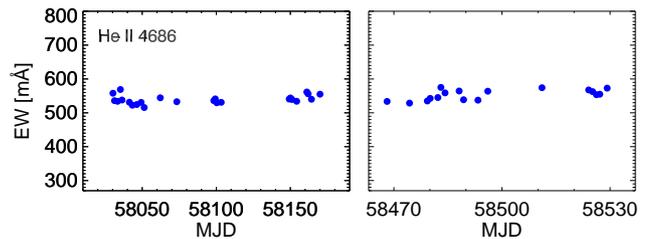}
                 \caption{
{\it Left:} EWs of the \ion{He}{ii}~4686 line measured in the FORS\,2 spectra obtained from
2017 October~4 to 2018 February~21.
{\it Right:} EWs of the \ion{He}{ii}~4686 line measured in the FORS\,2 spectra obtained from 
2018 December~16 to 2019 February~15.
}
   \label{fig:EWprev}
 \end{figure}

On the left panel of Fig.~\ref{fig:EWprev} we present the equivalent widths (EWs) of the \ion{He}{ii}~4686 line
measured in the FORS\,2 spectra obtained from 2017 October~4 to 2018 February~21 and those
from 2018 December~16 to 2019 February~15 on the right side.
No significant changes in line intensities are detected during both observing runs.


\section{Studying FORS\,2 spectra with insufficient signal-to-noise ratio}
\label{sec:lowSNR}

\begin{figure*}
\centering 
        \includegraphics[width=0.80\textwidth]{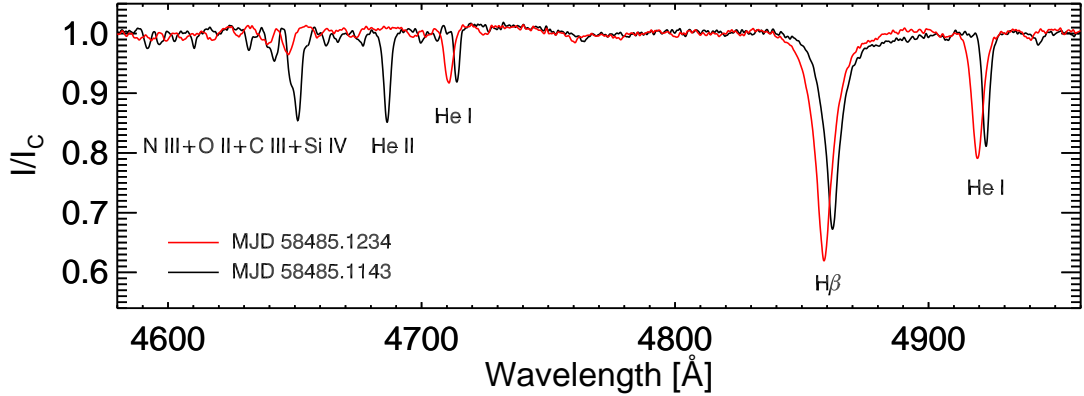}
               \caption{
Variability of line profiles detected in Stokes~$I$ spectra recorded on
the same night of 2019 January~1 and separated by only 25\,min. The spectra
were extracted using the MIDAS pipeline.
         }
   \label{fig:shift}
\end{figure*}

Rather unexpectedly, our analysis of the underexposed FORS\,2 spectra
from the observations between 2018 December and 2019 February showed that they 
are similar to the spectra obtained in the previously reported observations acquired with a $S/N$ of about 1130 
on 2018 February~17 \citep{Hubrig2019},
for which an increase of the
absolute value of the mean longitudinal magnetic field,
a change in spectral appearance,
and a decrease of the radial velocity by several 10\,km\,s$^{-1}$ were reported.
While such a behaviour was observed only once in the FORS\,2 observations 
acquired from 2017 October to 2018 February,  
it was detected in all five recent observations with a $S/N$ below 1300.
As an example, we present in Fig.~\ref{fig:shift} two FORS\,2 observations recorded with a $S/N$ of about 830 
on the same night on 2019 January~1 and separated by just 25\,min.
The first observation at MJD\,58485.1143 resulted in a $\left< B_{\rm z}\right>$ measurement
compatible with a non-detection, while the second observation at MJD\,58485.1234 gave
$\left< B_{\rm z} \right>=-1300\pm220$\,G, using all lines.
This spurious field increase at MJD\,58485.1234 is accompanied by a spurious change of 
spectral appearance, including the increase of all absorption hydrogen and \ion{He}{i} lines and the decrease of 
higher ionisation lines like \ion{He}{ii}, \ion{C}{iii} and \ion{Si}{iv}, and by a radial velocity shift of over 100\,km\,s$^{-1}$. 

\begin{figure}
\centering 
\includegraphics[width=0.80\columnwidth]{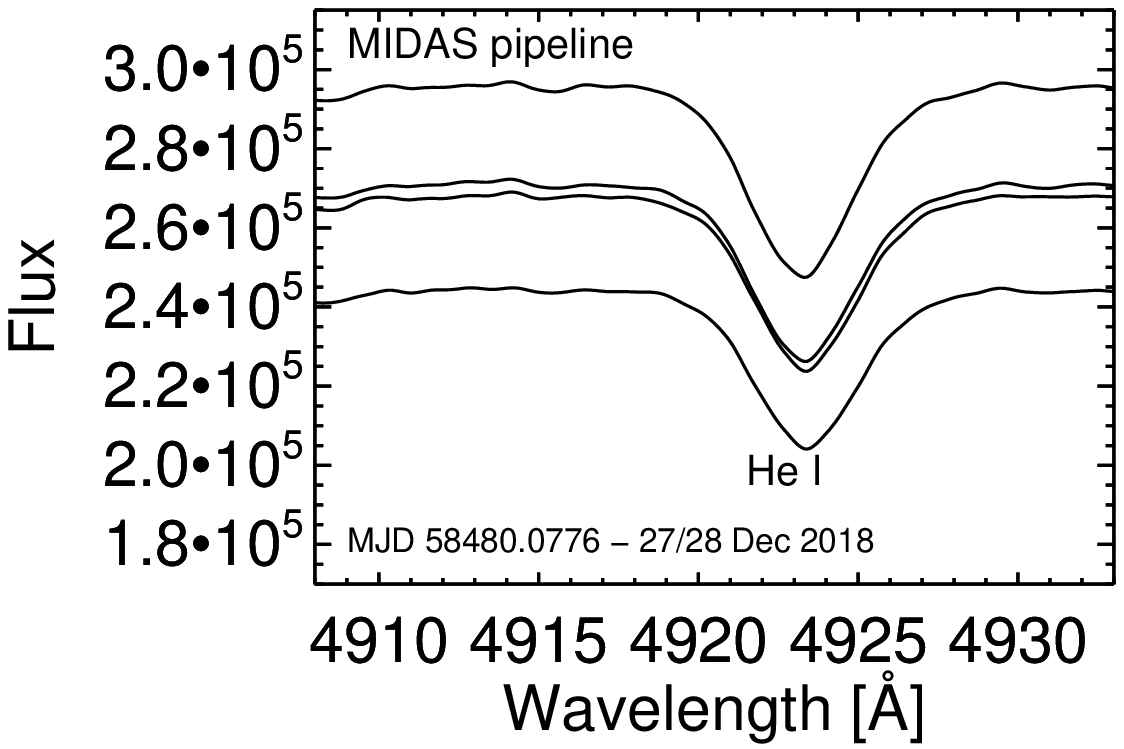}
\includegraphics[width=0.80\columnwidth]{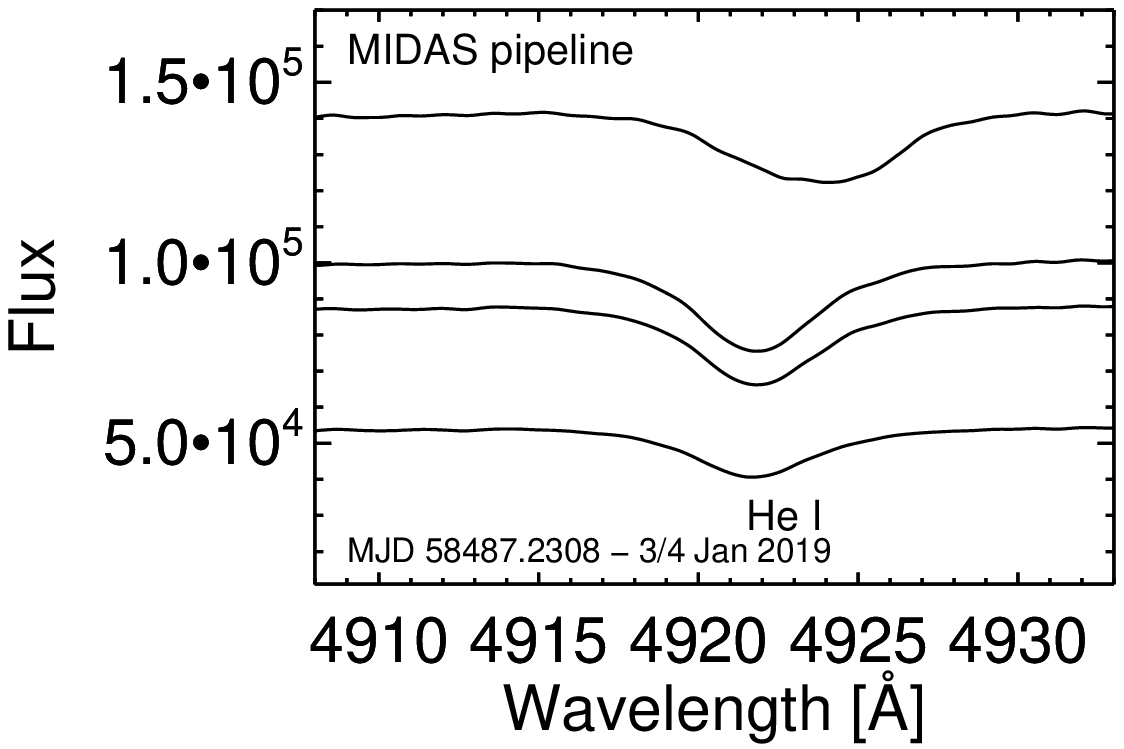}
               \caption{
Stokes~$I$ line profiles of \ion{He}{i}~4922 over the full
sequence of FORS\,2 subexposures obtained on a time scale of tens of seconds
at MJD\,58480.0776 ({\it top})
and on MJD\,58487.2308 ({\it bottom}).
While the extracted spectra at a higher $S/N$ of 2340 ({\it top}) do
not show a stable line profile, the extracted spectra at a lower
$S/N$ of 1278 ({\it bottom}) show significant wavelength jumps.
         }
   \label{fig:shift_6}
\end{figure}

In Fig.~\ref{fig:shift_6} we show the overplotted Stokes~$I$ spectra for \ion{He}{i}~4922 line profiles
corresponding to subexposures in observations of varying quality recorded on two different nights,
the observation obtained with a $S/N=2340$ at MJD\,58480.0776 and that with a $S/N=1120$ at MJD\,58487.2366.
According to the atmospheric parameters presented by \citet{Castro2015}, HD\,54879 has already
evolved from the ZAMS and is passing through the $\beta$~Cephei instability strip.
However, we are convinced that such short-term spectral variability is not real,
as we never detected it in previous observations of this and other targets
and it solely appears in data with insufficient $S/N$.

\begin{figure}
\centering 
\includegraphics[width=0.80\columnwidth]{shift_6_MIDAS.eps}
\includegraphics[width=0.80\columnwidth]{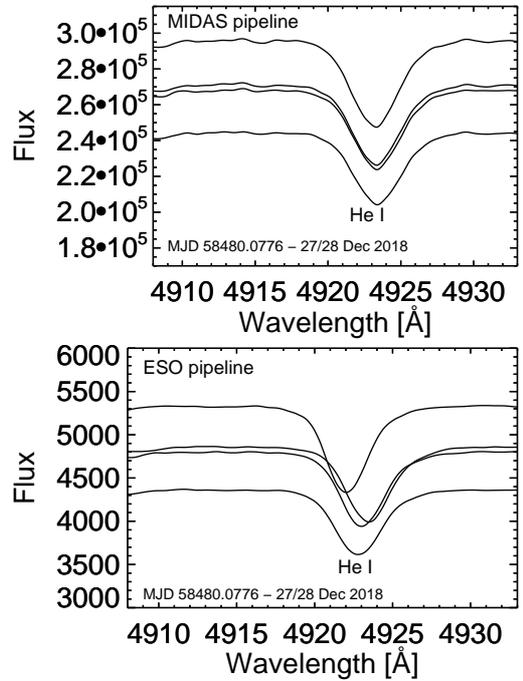}
               \caption{
Stokes~$I$ line profiles of \ion{He}{i}~4922 over the full
sequences of FORS\,2 subexposures  obtained on a time scale of tens of seconds at MJD\,58487.2366
at a $S/N$ of 2340.
The spectral extraction was carried out using the pipeline written
in the MIDAS environment ({\em top}) and the ESO FORS pipeline ({\em bottom}).
While the MIDAS pipeline shows no wavelength shifts between the different exposures,
the ESO FORS pipeline leads to wavelength shifts up the about 2\,\AA{}.
The difference in fluxes
between the two pipelines is caused by different ways in handling the absolute flux scale.
         }
   \label{fig:shift_11}
\end{figure}

Further, we investigated if an alternative spectrum extraction could result in more stable
wavelengths also for the spectra with insufficient $S/N$.
For this, we employed the ESO FORS pipeline based on the Reflex toolkit and tailored to
the polarimetric mode of FORS\,2 (PMOS).
However, as can be seen in Fig.~\ref{fig:shift_11},
the ESO FORS pipeline has issues with wavelength stability even for higher $S/N$ data.
While most of the higher $S/N$ spectra gave similar results when determining the longitudinal magnetic
field, when compared to the MIDAS pipeline results, we concluded that the ESO FORS PMOS pipeline
in its current form is not delivering proper results.

\begin{figure}
 \centering 
\includegraphics[width=0.80\columnwidth]{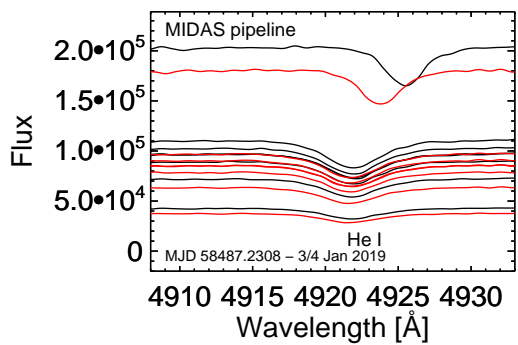}
                \caption{
Ordinary (red color) and extraordinary (black color)
circularly polarized line 
profiles of the \ion{He}{i}~4922 line extracted
using the MIDAS pipeline in observations at MJD~58487.2366. 
         }
   \label{fig:pmos}
\end{figure}

The observations with lower $S/N$ at MJD\,58487.2366 show a strong shift in the 
Stokes~$I$ profile recorded in one subexposure.
For the same night, we also present in Fig.~\ref{fig:pmos}
ordinary and extraordinary circularly polarized line profiles of the 
\ion{He}{i}~4922 line.
A clear shift in the spectra for one subexposure
indicates that the presence of a strong longitudinal magnetic field is spurious.

The inspection of all FORS\,2 extracted spectra
indicates no impact of high air mass or variable seeing during the observations.
The short-term spectral and magnetic variability is detected only in observations with a $S/N$ below 1300. 
We conclude that the spectral extraction
with the MIDAS pipeline is not working properly for underexposed spectra and is producing spectral artifacts.
While different scenarios were previously discussed by \citet{Hubrig2019} in an attempt to interpret the 
observation on 2018 February 17, it appears now that this observation was affected by an imperfect spectral extraction.

\section{Discussion} 
\label{sec:mod}

The new FORS\,2 spectropolarimetric data obtained from 2018 December to 2019 February
confirm the very slow magnetic field strength variability in HD\,54879.
While the few observations from 2014 and 2015 indicated a
mean longitudinal magnetic field value of the order of $-$500\,G to $-$900\,G, we observe in the last 
years a significantly weaker magnetic field with a mean longitudinal magnetic field value
between $-300$\,G and $+150$\,G. 
The strongest longitudinal magnetic field of positive polarity of 150\,G
was measured on the night of 2018 December~16.
After this date, the longitudinal magnetic field is gradually decreasing, reaching
a value of about $-$100\,G on 2019 February~15.
This slow magnetic field variability, related to the extremely slow rotation of HD\,54879, is also confirmed using high-resolution
HARPS and ESPaDOnS spectropolarimetry.
Assuming that the negative field extremum reaches a value of $-$500\,G to $-$900\,G,
measured in 2014 February,
the rotation cycle is expected to be longer than five years.  Additional evidence for a very
long rotation period, longer than nine years, follows from the consideration of the variability of the
H$\alpha$ line profiles. However, since very long rotation periods are best determined from magnetic field
variability, future monitoring of HD\,54879 should include both, the follow-up of the changes of the H$\alpha$
line profile and of the measurements of the longitudinal magnetic field. 

The analysis of the new FORS\,2 polarimetric spectra indicates that the
previous detection of a significant field increase and a change of the spectral appearance
is due to improper spectra extraction and wavelength calibration, using observations obtained
at an insufficient signal-to-noise ratio. 

Among the previously detected magnetic O-type stars, five are classified as Of?p stars. 
The primary characteristic for the Of?p stars, according to the definiton by \citet{Walborn1972},
is a variable and comparable emission strength of the C\,{\sc iii} blend
(C\,{\sc iii} $\lambda\lambda$4647-4650-4652) with respect to the
neighboring variable emission N\,{\sc iii} blend (N\,{\sc iii} $\lambda\lambda$4634-4640-4642).
The observed C\,{\sc iii} blend
in these stars is strongly variable, exhibiting transitions from absorption line profiles to emission 
line profiles at certain rotation phases. The presence of variable emission in the C\,{\sc iii} blend 
is indicative of circumstellar structure around the Of?p stars, related to their magnetospheres. 
However, the emission in the C\,{\sc iii} blend disappears entirely in late O-type stars
\citep{Walborn2010} and is thus missing in HD\,54879, meaning that a selective emission effect cannot
be observed in HD\,54879.
Obviously, the Of?p classification is very narrow as it is limited
to stars with spectral types in the range O4f?p--O8f?p,
with temperatures between 34.5\,kK and 41\,kK.
HD\,54879 is significantly cooler with $T_{\rm eff} = 30.5$\,kK and this is the main reason why
no association with the Of?p class has been done so far.

On the other hand, many properties of HD\,54879 are similar to those of Of?p stars.
The H$\alpha$ emission line in HD\,54879 presented in Fig.~\ref{fig:esp2} is highly variable. 
In analog to the C\,{\sc iii} blend, the presence of the H$\alpha$ emission in Of?p stars is related to 
their magnetospheres and its variability is expected to trace dense environments in Of?p stars.
This line is frequently used to determine their rotation periods. 

A remarkable resemblance of the ultraviolet (UV) spectra
of HD\,54879 and the Of?p star NGC\,1624-2 with a dipole strength of $\sim$20\,kG estimated by \citet{Wade2012}
was recently discussed by \citet{david2019}. The authors report that despite of the later spectral type
of HD\,54879, its UV spectrum is surprisingly similar to the UV spectrum of NGC\,1624-2 obtained
at a rotational phase of nearly magnetic equator-on view. Both stars exhibit the lowest $v \sin i$
and macroturbulent velocity $v_{\rm mac}$ values known among the magnetic O-type stars \citep{Shenar2017}
and do not show nitrogen excess \citep{Castro2015}.
Alike the spectral appearance of the Of?p star NGC\,1624-2, the high-resolution HARPS spectra
display weak emission lines belonging to various metal lines, indicating that the line formation in the atmosphere
of HD\,54879 can be similarly complex.

As we discussed in Sect.~\ref{sec:per}, the rotation period of HD\,54879 is probably longer than nine years.
Also Of?p stars are known as a class of slowly rotating magnetic massive stars with the
longest rotation period of about 50-60\,yr suggested for the Of?p star HD\,108 \citep{Naze2001}.
Furthermore, recent analyses of the {\em XMM-Newton} spectra of HD\,54879
by \citet{Shenar2017} and \citet{Hubrig2019}  indicate over-luminosity by at least one order compared to other
O-type stars with similar spectral types. Such an excess of X-ray luminosity is typical for
all Of?p stars, for which X-ray spectra are usually well  described  by  multi-temperature thermal plasma models.
All of these properties suggest that HD\,54879 is an analogue
to the Of?p stars and only misses the Of?p classification criteria because of its lower temperature.

In respect to the possible binary nature of HD\,54879, it was suggested in recent years that
three out of five Of?p stars are members of binary systems.
Long-term radial velocity changes indicating binarity were reported for the Of?p star
HD\,191612 \citep{Howarth2007} who suggested $P_{\rm orb}=1542$\,d.
According  to \citet{wade2019}, the Of?p star HD\,148937 is likely a high-mass,
double-lined spectroscopic binary \citep{wade2019}. 
Also for the Of?p star HD\,108 \citet{Naze2008} reported that a very long-term binary cannot be excluded.
Similarly, the compilation of radial velocity measurements over tens of years indicates that 
HD\,54879 could be a member of a long-period binary system.

The knowledge of the frequency of membership of upper main-sequence stars with radiative envelopes in wide binary systems 
 is very important, as it can be related to the origin of their magnetic fields.
It was suggested that magnetic fields may be generated by strong binary interaction, i.e.\
in stellar mergers, during a mass transfer, or in the course of a common envelope evolutionary phase
\citep{Tout2008}.
The resulting strong differential rotation \citep{Petrovic2005} is then considered as 
a key ingredient for the magnetic field generation.
Requiring mergers to produce magnetic stars implies that there should be almost no
magnetic star in a close binary. 
Indeed, magnetic components in close binaries are very rare:
Only three close binaries with a magnetic Ap component, HD\,98088 ($P_{\rm orb}=5.9$\,d; \citealt{Babcock1958}),
HD\,25267 ($P_{\rm orb}=5$\,d; \citealt{Borra1980}),
and  HD\,161701 ($P_{\rm orb}=12.5$\,d;  \citealt{Hubrig2014b}), are currently known, and only two late-B type binaries 
with magnetic  components  and orbital periods below 20\,d were recently detected, HD\,5550 
\citep{Alecian2016} and BD$-$19$^\circ$5044 \citep{Landstreet2017}.
The situation among early-B type stars is even more extreme, as only one early-B type short-period magnetic 
binary, HD\,136504  \citep{Shultz2015}, is currently known.
As wide binaries are widespread among Ap and late Bp stars \citep{Mathys2017}, such wide binaries 
could have been born as hierarchical triple stars, where the inner binary merged.

Observations in star-forming regions indicate that almost all stars form in clusters
\citep{Lada2003},
and that the number of multiple systems within these clusters is remarkably high. 
Binary population synthesis simulations predict that the rate of main-sequence mergers increases 
with mass (e.g., \citealt{mink2014,Schneider2015}).
\citet{mink2014} found a merger fraction of 8\% in a population of B-type stars and 12\% in O-type stars.
Also recent observations of Ap and late Bp stars support a scenario where 
mergers produce magnetic stars \citep{Mathys2017}. Clearly, future monitoring of radial velocities
of massive magnetic stars is important to be able to conclude on the role of binarity for the magnetic field generation.
  

\section*{Acknowledgments}
Based on observations made with ESO Telescopes at the La Silla Paranal
Observatory under programme IDs~191.D-0255, 0100.D-0110, and 0102.D-0285.
We thank the referee G.~Mathys for his constructive comments that helped to improve the paper.


\label{lastpage}


\begin{thebibliography}{}

\bibitem[\protect\citeauthoryear{Alecian et al.}{2016}]{Alecian2016}
  Alecian E., Tkachenko A., Neiner C., Folsom C.~P., Leroy B., 2016, A\&A, 589, A47 
  
\bibitem[\protect\citeauthoryear{Appenzeller et al.}{1998}]{Appenzeller1998} 
Appenzeller I., et al.,
1998, The ESO Messenger, 94, 1

\bibitem[\protect\citeauthoryear{Babcock}{1958}]{Babcock1958}
Babcock H. W., 1958, ApJS, 3, 14

\bibitem[\protect\citeauthoryear{Borra \& Landstreet}{1980}]{Borra1980} 
Borra E.~F., Landstreet J.~D., 1980, ApJS,  42, 421

\bibitem[\protect\citeauthoryear{Boyajian et al.}{2007}]{Boyajian2007}
Boyajian, T.~S., et al., 2007, PASP, 19, 742 

\bibitem[\protect\citeauthoryear{Castro et al.}{2015}]{Castro2015}
Castro N., et al.,
2015, A\&A, 581, A81

\bibitem[\protect\citeauthoryear{David-Uraz et al.}{2018}]{David2018}
David-Uraz A., Petit V., Erba C., Fullerton A., Walborn N., MacInnis R.,
2018, CoSka, 48, 134

\bibitem[\protect\citeauthoryear{David-Uraz et al.}{2019}]{david2019}
David-Uraz A., et al., 2019, MNRAS, 483, 2814  

\bibitem[\protect\citeauthoryear{de Mink et al.}{2014}]{mink2014}
de Mink S.~E., Sana H., Langer N., Izzard R.~G., Schneider F.~R.~N., 2014, ApJ, 782, 7

\bibitem[\protect\citeauthoryear{Donati et al.}{1997}]{Donati1997}
Donati J.-F., Semel M., Carter B.~D., Rees D.~E., Collier Cameron A., 1997,  MNRAS, 291, 658

\bibitem[\protect\citeauthoryear{Donati et al.}{2006}]{Donati2006}
Donati J.-F., Catala C., Landstreet J.~D., Petit P., 2006,
in Casini, R., Lites B.~W., eds, Astron.\ Soc.\ of the Pacific Conf.\ Ser., 358,
Solar Polarization 4, p.~362
  
\bibitem[\protect\citeauthoryear{Howarth et al.}{2007}]{Howarth2007}
Howarth I.~D., et al., 2007, MNRAS, 381, 433   

\bibitem[\protect\citeauthoryear{Hubrig et al.}{2014}]{Hubrig2014b}
Hubrig S. et al., 2014, MNRAS, 440, L6

\bibitem[\protect\citeauthoryear{Hubrig et al.}{2015}]{Hubrig2015}
Hubrig S., et al., 2015, MNRAS, 447, 1885

\bibitem[\protect\citeauthoryear{Hubrig et al.}{2019}]{Hubrig2019}
Hubrig S., et al., 2019, MNRAS, 484, 4495

\bibitem[\protect\citeauthoryear{J\"arvinen et al.}{2017}]{Jarvinen2017}
J\"arvinen S.~P., Hubrig S., Ilyin I., Shenar T., Sch\"oller M.,
2017, Astron.\ Nachr., 338, 952

\bibitem[\protect\citeauthoryear{Kupka et al.}{2011}]{kupka2011}
Kupka F., Dubernet M.-L., VAMDC Collaboration, 2011, BaltA, 20, 503 

\bibitem[\protect\citeauthoryear{Lada \& Lada}{2003}]{Lada2003}
Lada C.~J., Lada E.~A., 2003, ARA\&A, 41, 57

\bibitem[\protect\citeauthoryear{Landstreet et al.}{2017}]{Landstreet2017}
Landstreet J.~D., et al., 2017, A\&A, 601, A129

\bibitem[\protect\citeauthoryear{Martins et al.}{2012}]{Martins2012}
Martins F., Escolano C., Wade G.~A., Donati J.~F., Bouret J.~C., Mimes Collaboration,
2012, A\&A, 538, 29

\bibitem[\protect\citeauthoryear{Mathys}{2017}]{Mathys2017}
Mathys G., 2017, A\&A, 601, A14

\bibitem[\protect\citeauthoryear{Naze et al.}{2001}]{Naze2001}
Naz\'e Y., Vreux J.-M., Rauw G., 2001, A\&A, 372, 195 

\bibitem[\protect\citeauthoryear{Naze et al.}{2008}]{Naze2008}
Naz\'e Y., Walborn N.~R., Martins F., 2008, RMxAA, 44, 331  

\bibitem[\protect\citeauthoryear{Neubauer}{1943}]{Neubauer1943}
Neubauer F.~J., 1943, ApJ, 97, 30

\bibitem[\protect\citeauthoryear{Petrovic et al.}{2005}]{Petrovic2005}
Petrovic J., Langer N., Yoon S.-C., Heger A., 2005, A\&A, 435, 247

\bibitem[\protect\citeauthoryear{Schneider et al.}{2015}]{Schneider2015}
Schneider F.~R.~N., Izzard R.~G., Langer N., de Mink S.~E., 2015, ApJ, 805, 20

\bibitem[\protect\citeauthoryear{Shultz et al.}{2015}]{Shultz2015}
Shultz M., Wade G.~A., Alecian E., BinaMIcS Collaboration, 2015, MNRAS, 454, L1

\bibitem[\protect\citeauthoryear{Sundqvist et al.}{2012}]{Sundqvist2012}
Sundqvist J.~O., ud-Doula A., Owocki, S.~P., Townsend R.~H.~D., Howarth I.~D., Wade G.~A., 2012, MNRAS, 423, L21

\bibitem[\protect\citeauthoryear{Shenar et al.}{2017}]{Shenar2017}
Shenar T., et al.,
2017, A\&A, 606, A91

\bibitem[\protect\citeauthoryear{Shultz \& Wade}{2017}]{Shultz2017}
Shultz M., Wade G.~A., 2017, MNRAS, 468, 3985 

\bibitem[\protect\citeauthoryear{Snik et al.}{2008}]{Snik2008}
  Snik F., Jeffers S., Keller C., Piskunov N., Kochukhov O., Valenti J., Johns-Krull C., 2008,  in McLean I. S.,
  Casali M. M., eds, Proc.\ SPIE Conf.\ Ser.\ Vol.~7014,
  Ground-based and Airborne Instrumentation for Astronomy~II.
  SPIE, Bellingham, p.~E22

\bibitem[\protect\citeauthoryear{Tout et al.}{2008}]{Tout2008}
Tout C.~A., Wickramasinghe D.~T., Liebert J., Ferrario L., Pringle J.~E.,
2008, MNRAS, 387, 897	

\bibitem[\protect\citeauthoryear{Wade et al.}{2012}]{Wade2012}
Wade G.~A., et al., 2012, MNRAS, 425, 1278

\bibitem[\protect\citeauthoryear{Wade et al.}{2019}]{wade2019}
Wade G.~A., et al., 2019, MNRAS, 483, 2581 

\bibitem[\protect\citeauthoryear{Walborn}{1972}]{Walborn1972}
Walborn N.~R., 1972, AJ, 77, 312

\bibitem[\protect\citeauthoryear{Walborn et al.}{2010}]{Walborn2010}
Walborn N.~R., et al.,  2010, ApJ, 711, L143

\end{thebibliography}
\end{document}